\documentclass[10pt,twoside]{article}
\usepackage[english]{babel}

\usepackage {graphicx}
\usepackage {graphics}
\usepackage {floatflt}
\usepackage {latexsym}
\usepackage {caption3} 
\usepackage[labelsep=period]{caption}[2007/11/04 v.3.1e]
\begin{document}

{\begin{center}
{\underline{\scriptsize{Journal of Siberian Federal University. Mathematics \& Physics 2011, 4(1), 3--10}}}
\end{center}}

\vspace{0.5cm}
{\raggedright
UDK 531.51}

\vspace{0.5cm}
{\raggedright
{\bf\Large Lightons and Helixons as Lightlike Particles \\} 
{\raggedright{\bf\Large in General Relativity}}

\vspace{0.5cm}

{\raggedleft{\large \bfseries Alexandre~M.Baranov~\footnote{alex\_m\_bar@mail.ru; AMBaranov@sfu-kras.ru}}

{\raggedleft\small\mdseries Institute of the Engineering Physics
\& Radio-Electronics,\\
Siberian Federal University{\footnote{\copyright ~Siberian Federal University. All rights reserved}},\\
Svobodny 79, Krasnoyarsk, 660041\\
Russia\\}}}

{\begin{center}
{\underline{\scriptsize{Received 07.07.2010, received in revised form 10.09.2010, accepted 20.10.2010}}}
\end{center}}

{{\small{\it The exterior gravitational fields of the massive Schwarzschild, Kerr and NUT particles have an algebraic type {\bfseries D} according to Petrov's algebraic classification of the gravitational fields. A lightlike limit is input for these particles. It is shown that under this limiting procedure the gravitational fields of the particles are transformed into the gravitational fields of the lightlike particles a lighton and a helixon. On the other hand this limit can be described as a cusp catastrophe on a level of Weyl's matrix with a change of a gravitational field symmetry of such sources. In considered cases we have a phase gravitational transition of second kind from {\bfseries D} type into {\bfseries N} type or {\bfseries III} type. There is a transition of one "phase" to another. Petrov's algebraic types are the different "phases" of the gravitational field. It is shown that the lightlike sources in General Relativity "have no hairs".

}}}

\smallskip

\vspace{0.5cm}
A task of the finding a field of an electric charge when its velocity tends to the velocity of light is known in the classical electrodynamics \cite{dau1}. A limiting field of such fast moving charge particle approaches to the field of a monochromatic electromagnetic plane wave. In General Relativity (GR) there is a similar problem for some metrics of the particlelike solutions of Einstein's gravitational equations.

The exterior metrics of the gravitational fields of the Schwarzschild, Kerr and NUT solutions are well-known. These solutions describe the gravitational fields of massive particles. In an article a lightlike limiting procedure of the massive particles is introduced in \cite{bam1,bam2}. A requirement of a particle velocity to tend to the velocity of light is applied to these metrics. A correct finding of the limit expressions is connected with an using of the generalized singular functions ($\delta$-functions of Dirac) \cite{bam1}. A demonstration of such correct using of the $\delta$-functions to the task of a rapidly moving charge in the electrodynamics is in \cite{bam1,bam2}.

In General Relativity the limit particle velocity $V$ tends to the velocity of light $c$ (here we have $V \to c= 1$), and a rest mass of the particle tends to zero ( $m_0 \to 0$) so that a total relativistic energy of the particle is a constant, $E = const$. We shall call this procedure as a {\it lightlike limit} (see for example \cite{bam1,bam6}).

An aimpoint of this paper is to summarize the results of the lightlike limits for the solutions of Schwarzschild, Kerr and NUT on a level of an algebraic classification of Petrov and to find the algebraic types of the limiting gravitational fields. Also we must investigate a possibility of a superposition of the Einstein equations' limiting solutions.

\section{Lightlike limit of Schwarzschild metric}

An exterior gravitational field of a rest massive particle is described by the Schwarzschild solution \cite{syng1}. A metric of this solution can be written in the Kerr-Schild form

\begin{equation}
{\rm  }g_{\mu \nu } = \delta _{\mu \nu } - 2Hl_\mu l_\nu  ,
\label{eq:1}
\end{equation}

\noindent
where $\mu ,\nu = 0,1,2,3;$ $\delta _{\mu \nu } = diag(1, - 1, - 1, - 1);$ $ H = m_0 /r ;$ $l_\mu = (t + r + )_{,\mu };$ $l_\mu  l^\mu = 0;$ $m_0$ is the rest mass; $r$ is a radial variable; the gravitational constant of Newton is an unit here; a comma denotes a partial derivative. Now we shall consider the lightlike limit for the Schwarzschild-like metric (\ref{eq:1}). The square of an interval with the metric tensor (\ref{eq:1})

\begin{equation}
ds^2 = ds_0^2 - 2 H l_\mu l_\nu d x^\mu d x^\nu 
\label{eq:2}
\end{equation}

\noindent
(with $ds_0^2 =\delta _{\mu \nu } d x^\mu d x^\nu  $) is an invariant under the Lorentz transformations

\begin{equation}
t \to (t + zV)/(1 - V^2 )^{1/2}; \;\; z \to (z + Vt)/(1 - V^2 )^{1/2}; \;\; x \to x; \;\; y \to y.
\label{eq:3}
\end{equation}

After applying these transformations to the metric with (\ref{eq:1}) we can rewrite the interval as

\begin{equation}
ds^2  = ds_0^2  - 2\tilde H\tilde l_\mu  \tilde l_\nu  d\tilde x^\mu  d\tilde x^\nu ,
\label{eq:4}
\end{equation}

\noindent
where $\tilde H = E\left((z + Vt)^2 + (1 - V^2 )(x^2 + y^2 )\right)^{-1/2},$ $\tilde l_{\mu} ,$ $d\tilde x^\mu $ are new values after the transformations; $m_0 = E(1 - V^2 )^{1/2};$ $E$ is a total energy of the particle and here $E = const.$ 

We must note here that the limits of the functions when $V \to 1$ can be found both in a class of the usual functions and in a class of the generalized functions. Let's consider one example.

If for any function $f = f(\xi ,V)$ with some variable $\xi  $ and parameter $V$ takes place 

\begin{equation}
\mathop {\lim }\limits_{V \to 1} \left( \int\limits_{ - \infty }^\infty {f(\xi ,V)d\xi } \right) = b = const \ne 0,
\label{eq:5}
\end{equation}
\noindent
then in the class of the generalized functions $\mathop {\lim }\limits_{V \to 1} f(\xi ,V) = b \cdot \delta (\xi )
$ with $\delta $-function of Dirac. 

The other examples is the limits of the functions $f_1 = \varepsilon ^2 (\xi ^2 + \varepsilon ^2 \rho ^2 )^{ - 3/2}$ and 
$f_1  = (\xi ^2  + \varepsilon ^2 \rho ^2 )^{ - 1/2}$ when $\varepsilon  \to 0$ with $(1 - V^2 ) \equiv \varepsilon ^2;$ $x^2  + y^2  \equiv \rho ^2 ;$ $z + Vt \equiv \xi;$ $z + t \equiv v.$ The limit of an integral $\mathop {\lim }\limits_{\varepsilon  \to 1} \int\limits_{ - \infty }^{\infty `} {f_1 (\xi ,\varepsilon )d\xi }$ is equal to 
$2/\rho ^2.$ Therefore the complete limit of the function $f_1 (\xi ,\varepsilon )$ is 
\begin{equation}
\mathop {\lim }\limits_{\varepsilon  \to 1} ( \varepsilon ^2 (\xi ^2  + \varepsilon ^2 )^{ - 3/2} )  = (2/\rho ^2 )\cdot\delta (v) = (2/(x^2  + y^2 ))\cdot\delta (z + t).
\label{eq:6}
\end{equation}

Here an usual limit of the function $f_1 (\xi ,\varepsilon )$ equals zero. 
The limit of the function $f_2 (\xi ,\varepsilon )$ when $\varepsilon  \to 0$ is equal to $|1/|v|$ 
in the class of the usual functions and $\mathop {\lim }\limits_{\varepsilon  \to 1} \int\limits_{ - \infty }^{\infty `} {f_2 (\xi ,\varepsilon )d\xi }  =  - 2 \cdot \ln (\rho ).$ Thus we have

\begin{equation}
\mathop {\lim }\limits_{\varepsilon  \to 0} (\xi ^2 + \varepsilon ^2 \rho ^2 )^{ - 1/2} =
 |v|^{ - 1} - 2(\ln \rho )\delta (v).
\label{eq:7}
\end{equation}

Next limits are correct in the class of the generalized functions for the modified metric (\ref{eq:3}) with

\begin{equation}
\mathop {\lim }\limits_{V \to 1} \tilde l_0 = 1;
\label{eq:8}
\end{equation}

\begin{equation}
\begin{array}{crc}
\mathop {\lim }\limits_{\varepsilon  \to 1} \{ \tilde H\tilde l_1^2 \varepsilon ^2 \}  = \mathop {\lim }\limits_{\varepsilon  \to 1} \{ \tilde H\tilde l_2 \varepsilon \}  = \mathop {\lim }\limits_{\varepsilon  \to 1} \{ \tilde H\tilde l_2^2 \varepsilon ^2 \}  = \mathop {\lim }\limits_{\varepsilon  \to 1} \{ \tilde H\tilde l_1 \varepsilon \} = \mathop {\lim }\limits_{\varepsilon  \to 1} \{ \tilde H\tilde l_3 \tilde l_1 \varepsilon \} = \\
\\ 
 = \mathop {\lim }\limits_{\varepsilon  \to 1} \{ \tilde H\tilde l_3 \tilde l_2 \varepsilon \}  = \mathop {\lim }\limits_{\varepsilon  \to 1} \{ \tilde H\tilde l_2 \tilde l_1 \varepsilon \} = 0;
\end{array}
\label{eq:9}
\end{equation}

\begin{equation}
H_0  = \mathop {\lim }\limits_{V \to 1} \{ \tilde H\}  = \mathop {\lim }\limits_{V \to 1} \{ \tilde H\tilde l_3^2 \}  = \mathop {\lim }\limits_{V \to 1} \{ \tilde H\tilde l_3 \}  = E(|v|^{ - 1}  - 2\delta (v)\ln \rho ).
\label{eq:10}
\end{equation}

The limiting metric obtained from (\ref{eq:4}) is also Kerr-Schild's metric with another the lightlike vector $k_\mu $ \cite{bam1,bam4},
\begin{equation}
g_{\mu \nu }  = \delta _{\mu \nu }  - 8H_0 k_\mu  k_\nu , 
\label{eq:11}
\end{equation}
\noindent
where $k_\mu = \delta _\mu ^0 + \delta _\mu ^3;\;$ $k_\mu  k^\mu = 0;\;$ $\det (g_{\mu \nu } ) = \det (\delta _{\mu \nu } ) = - 1$ (see \cite{bam7}). 

The same limiting metric (\ref{eq:11}) was obtained in \cite{AS1} for the exterior metric of Schwarzschild, but in the homogeneous coordinates.

The Einstein equations in such limiting case are the wave equation with a singular energy-momentum tensor $T_{\mu \nu },$ which describes lightlike radiation,
\begin{equation}
\Box {g_{\mu \nu }} = - 8\pi T_{\mu \nu } = - 8\pi \{ 2E\delta (z + t)\delta (x)\delta (y)k_\mu k_\nu \}, 
\label{eq:12}
\end{equation}
\noindent
where $\Box$ is D'Alamber's operator in the Minkowski space-time.

Therefore here we have a singular scalar particle moving with the light velocity and having zero rest mass. We will name this particle as a {\it lighton}.

The integral limit (\ref{eq:7}) is an invariant under a substitution $\rho  \to \alpha \rho $ with 
$\alpha = const$ and $[\alpha ] = \hbox{1/cm}.$ Thus we can rewrite (\ref{eq:10}) as 
\begin{equation}
H_0 = E(|z + t|^{ - 1}  - 2\delta (z + t)\ln (\alpha (x^2  + y^2 )^{1/2} )).
\label{eq:13}
\end{equation}
Now the metric (\ref{eq:11}) can be rewritten if we will use the retard and advance time variables together with the polar coordinates after the tranformation $H_0 \to H_0 /4\;$ as 
\begin{equation}
ds^2  = dudv - d\rho ^2  - \rho ^2 d\varphi ^2 - 2H_0 dv^2 ,
\label{eq:14}
\end{equation}
\noindent
where $\varphi$ is an angular variable. 

A singularity $1/|v|$ in the function $H_0$ can be excluded by the coordinate transformation
\begin{equation}
u \to u + 2E\ln |v|;\;\;\;v \to v
\label{eq:15}
\end{equation}
\noindent
and we can introduce a new function
\begin{equation}
H_0 \to H_0 = - E\delta (v)\ln (\alpha \rho ).
\label{eq:16}
\end{equation}
\noindent
The condition $\alpha \rho  = 1$ sets a horizon as a circle where $H_0 \equiv 0.$ It is a consequence 
of an existence of Schwazschid's horizon.

The metric (\ref{eq:11}) has two Killing's vectors: a {\it lightlike vector} 
$\xi _L  = (\partial /\partial t + \partial /\partial z)$ and a {\it spacelike vector} 
$\xi _Z  = (x\partial /\partial y - y\partial /\partial x),$ which defines the axial symmetry and in the polar coordinates it equals to $\partial /\partial \varphi .$ The Schwarschild-like solution has four Killing's vectors: a {\it timelike vector} $\xi_T = \partial /\partial t $ and {\it three spacelike vectors}: 
$\xi_X  = (y\partial /\partial z - z\partial /\partial y);\;$ 
$\xi_Y  = (z\partial /\partial x - x\partial /\partial z);\;$ 
$\xi_Z  = (x\partial /\partial y - y\partial /\partial x).$ 
Further the Lorentz boost is applied to Killing's vectors of Schwarzschild's solution with the lightlike limit. 
The vector $\xi_Z $ will be kept invariable. Vectors $\mathord{\buildrel{\lower3pt\hbox{$\scriptscriptstyle\frown$}} 
\over L} \xi_T ,\mathord{\buildrel{\lower3pt\hbox{$\scriptscriptstyle\frown$}} 
\over L} \xi_X ,\mathord{\buildrel{\lower3pt\hbox{$\scriptscriptstyle\frown$}} 
\over L} \xi_Y $ will be degenerated into the lightlike vector $\xi _L .$

Thus the lightlike limit of such massive particle leads to a change of an exterior gravitational field symmetry and to an appearance of a new scalar particle: a {\it lighton}.

The gravitational field of the lighton belongs to the algebraic type of {\it{\bfseries N}} (a wave type) \cite{bam1,bam2,bam5}.

\section{Lightlike limit of Kerr metric}

The metric of Kerr's solution \cite{ker1} describing the exterior gravitational field of the rest massive particle with an angular momentum can be written in the Kerr-Schild form (\ref{eq:1}) with 
\begin{equation}
H = (m_0 /2)(1/\omega  + 1/\omega ^* );\omega ^2  = x^2  + y^2  + (z - ia)^2 ;
\label{eq:17}
\end{equation}
\begin{equation}
l_0 = 1;\;\;\;l_k  = (\omega _{,k}  + \omega _{,k}^*  - i\varepsilon _{kmn} \omega _{,m} \omega _{,n}^* )/(1 + \omega _{,n} \omega _{,n}^* ),
\label{eq:18}
\end{equation}
\noindent
where the roman indexes $k,n,m = 1,2,3;$ $i^2 = - 1;$ a star symbol $*$ is an operation of a complex conjugation; $a = M_z /m_0;$ $M_z $ is a proper angular momentum (a spin) of the particle  along of $z$ axis; $\varepsilon _{ijk}$ is Levi-Civita's symbol. If the energy $E$ is the constant under the Lorentz transformation, then $M_z = \tilde M_z$ and $a(1 - V^2 ) \equiv J = const.$ 

For a case of a slow rotation of Kerr's source $J/E <  < 1$ the lightlike limit leads to the limiting metric \cite{bam1,bam4,bam4a,bam8}
\begin{equation}
g_{\mu \nu } = \delta _{\mu \nu } - Q_{\mu \nu }, 
\label{eq:19}
\end{equation}
\noindent
where 
\begin{equation}
Q_{\mu \nu }  = 8H_0 k_\mu  k_\nu   - 8(\partial H_0 /\partial y)Jn_{(\mu } k_{\nu )}  + 8(\partial H_0 /\partial x)Jm_{(\mu } k_{\nu )}; 
\label{eq:20}
\end{equation}
\noindent
$n_\mu = \delta _\mu ^1;$ $m_\mu   = \delta _\mu ^2 ;\;$ $Q_{\mu \nu }$ is a nilpotent tensor of an index three 
and $Sp\, Q^2  = 0.$ Then $\det (g_{\mu \nu } ) = - 1$ (see \cite{bam7}). 

In this limiting case the Einstein equations are the linear wave equations with the energy-momentum tensor 
of {\it "a rotational lightlike source"} (after a scale transition $E \to E/4;J \to 2J$)
\begin{equation}
\Box Q_{\mu \nu } = - 16\pi T_{\mu \nu },
\label{eq:21}
\end{equation}
\vspace{-0.2cm}
\noindent
where 
\begin{equation}
T_{\mu \nu } = T_{00} (k_\mu  k_\nu + (J/y)n_{(\mu } k_{\nu )} - (J/x)m_{(\mu } k_{\nu )} ).
\label{eq:22}
\end{equation}
\noindent
Here we used $\partial \delta (x)/\partial x =  - \delta (x)/x$ and 
$T_{00}  = E\delta (z + t)\delta (x)\delta (y).$

Thus we obtained a singular lightlike particle with a helicity $M_Z  =  \pm JE$ as 
a photon \cite{bam1,bam4,bam4a, bam8}. 
We will name this particle as {\it a helixon}.

The lightlike limiting metric (\ref{eq:19}) has two Killing's vectors only: {\it a lightlike vector} 
$\xi _L = (\partial /\partial t + \partial /\partial z)$ and {\it an axial spacelike vector} $\xi _Z = \partial /\partial \varphi $ in polar coordinates. The Kerr solution has two Killing vectors: the timelike vector 
$\xi _T = \partial /\partial t$ and {\it the axial spacelike vector} 
$\xi _Z = \partial /\partial \varphi .$

The Lorentz boost is applied to Killing's vectors of Kerr's solution together with the lightlike procedure and leads to $\xi _Z  \to \xi _Z $ ; $\xi _Z  \to \xi _L .$

Hence the lightlike limit of Kerr's massive particle modifies the exterior gravitational field symmetry and leads to the appearance of the new spinning lightlike particle: {\it the helixon.}

When a vector of the relative angular momentum ($\vec{a} = (M_Z /m_0)\, \vec{n} \;;\; \vec{n}^2 =1$) is a perpendicular to $z$ axis then the lightlike limit procedure leads to a loss of the limiting value of Kerr's relative angular momentum. In this case under the lightlike limit we have the same result as for Schwarzschild's solution and the lightlike limit's particle is the scalar particle ({\it the lighton}) \cite{bam4,bam4a,bam8}.

The gravitational field of the helixon belongs to the algebraic wave type of {\it {\bfseries III}} with a lengthwise component of the gravitational field \cite{bam1, bam2,bam5}

\section{Lightlike limit of NUT metric}

The exterior gravitational field of the massive particle describing by NUT's solution \cite{NUT1,misn1} has the algebraic type {\it {\bfseries D}} with a dual mass $b$ (as a gravitational analog of a dual magnetic charge). 

Under the lightlike limit when the particle velocity tends to the velocity of light along $z$  axis, the total energy $E = m/\varepsilon $ of the particle is the constant (i.e. $m_0  \to 0$ , when $V \to 1$) and NUT's parameter $b \to 0$ so that the new parameter $B = b/\varepsilon $ is equal to the constant \cite{bam3,bam9}.

We can write the limiting metric in the polar coordinates as a result of the lightlike limit
\begin{equation}
ds^2 = dudv - d\rho ^2 - \rho ^2 d\varphi ^2  - 4H_0 dv^2  - 4Bdvd\varphi ,
\label{eq:23}
\end{equation}
\noindent
where $u = t - z;$ $v = t + z;$ $\rho $ and $\varphi $ are polar coordinates; $H_0 $ is given by (\ref{eq:15}).

Nevertheless under the coordinate transformations 
\vspace{-0.2cm}
\begin{equation}
u \to u = 4B\varphi ;v \to v
\label{eq:24}
\end{equation}
\noindent
the metric (\ref{eq:23}) is transformed to the lighton  metric (\ref{eq:13}). 

Hence the NUT parameter in the lightlike limit is lost and we obtain the lightlike limit of NUT particle as the {\it lighton}.

The gravitational field of the lightlike limit of NUT particle belongs to the algebraic type of {\it{\bfseries N }}
(the wave type) \cite{bam3,bam9} also as the lightlike limit of Schwarzschid particle.

\section{On construction of lightlike sources}

As well known a focusing effect of the lightlike geodesic lines is absent if all the lightlike particles rapidly move in the same direction. In this case we can construct some lightlike sources because the Einstein equations are linear.

At first we will consider a lightlike plane which is perpendicular to $z$ axis. After an integration of an energy density $T_{00} = E \delta (x)\delta (y)\delta (z + t)$ of a lightlike source in a plane of the variables $x$ and $y$ we obtain a new energy density which equals to $T_{00} = E\delta (z + t).$ The energy-momentum tensor will be equal to $T_{\mu \nu }  = E\delta (z + t)k_\mu  k_\nu $ and a new function $H_0 $ will  be equal to
\begin{equation}
H_0  =  - 2\pi E(x^2  + y^2 )\delta (z + t) =  - 2\pi E\rho ^2 \delta (v).
\label{eq:25}
\end{equation}
A summarized gravitational field is the gravitational field of a lightlike plane front consisting of the continuum of the {\it lightons}. Such gravitational field belongs to the algebraic type {\it{\bfseries 0}} (i.e. the conformal flat type).

The other case is when we will integrate over all  $3D$ the spatial variables $x,$ $y,$ $z.$ The energy-momentum tensor $T_{\mu \nu }$ and the function $H_0 $ will be accordingly written as 
\begin{equation}
T_{\mu \nu } = Ek_\mu  k_\nu ; \;\; H_0 =  - 2\pi E\rho ^2 .
\label{eq:26}
\end{equation}
The gravitational field of such lightlike medium belongs to the algebraic type {\it{\bfseries 0}}.

In third case we can construct the lightlike pencil (a {\it photon hairline}) as a superposition of the singular lightlike sources \cite{bam4,bam4a,bam10,bam11}. Thus for monochromatic lightlike particles with the same total energy $E$ and the helicity $M_Z = J \cdot E$ we will make metric in the cylindrical coordinates for an infinite lightlike pencil as
\begin{equation}
ds^2 = dudv - d\rho ^2 - \rho ^2 d\varphi ^2 - 2Hdv^2 - 8JEdvd\varphi, 
\label{eq:27}
\end{equation}
\noindent
where $H =  - 4E\ln (\alpha \rho ).$

This metric is reduced to
\begin{equation}
ds^2  = dudv - d\rho ^2  - \rho ^2 d\varphi ^2  - 2Hdv^2 
\label{eq:28}
\end{equation}
\noindent
by the coordinate transformation $u \to u - 8M_Z \varphi.$ Therefore the infinite monochromatic lightlike pencil has not a constant spin (the {\it helicity}).

However, if we have a monochromatic lightlike ray with an endpoint which equals not the infinity then such ray can have the helicity for the function
\begin{equation}
H = - 4E \cdot \Theta (v)\ln (\alpha \rho )
\label{eq:29}
\end{equation}
\noindent
with a step function 
\vspace{-0.4cm}
\begin{equation}
\Theta (v) = \Theta (t + z) = \int\limits_{ - \infty }^{t + z} {\delta (\tau )d\tau }, 
\label{eq:30}
\end{equation}
\noindent
where $\delta (v)$ is Dirac's singular function.

Hence as the result of the lightlike procedure of Schwarzschild's, NUT's and Kerr's particle-like solutions we obtain the new metrics of the gravitational wave fields of the lightlike massless particles ({\it lightons} and {\it helixons}) with two freedom parameters only: the total energy and the helicity. The another physical parameters are lost under such limiting process. We can say that the lightlike sources in General Relativity {\it "have no hairs"} \cite{bam4,bam4a}.

\section{Lightlike limit and theory of catastrophe}

\begin{figure}[t,h]
\begin{center}
\fbox{\parbox{6.5cm}{\rule[-0.5cm]{0 mm}%
{6.5cm}\hfil\centering\includegraphics[width=6.5cm]{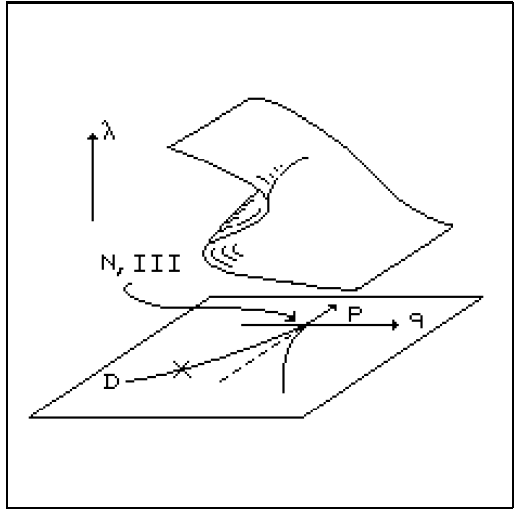}}}
\caption{The cusp catastrophe's surface and its projection onto the plane of the control parameters $p$ and $q$.}
\end{center}
\end{figure}

From the point of view of a catastrophe theory such lightlike limit is a {\bf catastrophe} on the level of Weyl's matrixes. These matrixes can be constructed by way of the mapping 
the Weyl tensor 
\vspace{-0.3cm}
\begin{eqnarray} 
W_{\alpha \beta \gamma \delta}= R_{\alpha\beta\gamma\delta} + R_{\gamma[\alpha}g_{\beta]\delta}  - R_{\delta[\alpha}g_{\beta]\gamma} -\frac{R}{3} g_{\gamma[\alpha} g_{\beta]\delta}
\label{eq:31}
\end{eqnarray}
\noindent
into the traceless $3\times3$ Weyl matrix 
\vspace{-0.2cm}
\begin{equation} 
{\bf \hat{W}} = (W_{kj}) = (\Omega_k^{\alpha \beta} \Omega_j^{\gamma \delta} W_{\alpha \beta \gamma \delta})
\label{eq:32}
\end{equation}
\noindent
with the help of a projector 
\vspace{-0.2cm}
\begin{equation} 
\Omega_k^{\alpha \beta} =\delta_{ [k }^\alpha \delta_{ 0] }^\beta - \frac{i}{2} \varepsilon_{kmn} \delta_m^\alpha \delta_n^\beta,
\label{eq:33}
\end{equation}
\noindent
where $i^2 = -1;\;$ $\delta_m^\alpha$ is the Kronecker symbol, the square brackets note an antisymmetric operation.

Now we shall consider an eigenvalue problem
\begin{equation} 
{\bf \hat{W}}{\bf X} = \lambda {\bf X},
\label{eq:34}
\end{equation}
\noindent
where ${\bf X}$ is an eigenvalue vector and $\lambda$ is an eigenvalue value. A characteristic equation is 
\begin{equation} 
det({\bf \hat{W}}-\lambda {\bf \hat{I}}) \sim\lambda^3 +p\lambda +q = 0\,
\label{eq:35}
\end{equation}
\noindent
with control parameters $\,p = p(\varepsilon),$ $q = q(\varepsilon)$ and a potential function 
\begin{equation} 
V=(1/4){\lambda}^4+(1/2){\lambda}^2 p+{\lambda}q,
\label{eq:36}
\end{equation}
\noindent
i.e. we have a cusp catastrophe. Here ${\bf \hat{I}}$ is a unit matrix, ${\bf \hat{I}} = diag (1,1,1).$

A discriminant of the equation (\ref{eq:34}) is $Q=(p/3)^3+(q/2)^2.$ When $Q=0$ we have a semicubical parabola $p=-3(q/2)^{2/3}$ which corresponds to Weyl's matrix of {\it{\bfseries D}} type (see Fig.1). Our case is marked by a cross ($q < 0$).

When $V \rightarrow 1$ (the lightlike limit) then $\varepsilon \rightarrow 0 $
 and $p(\varepsilon) \rightarrow 0,\; q(\varepsilon ) \rightarrow 0 $. 
As we earlier saw the gravitational field symmetry is lost under such lightlike limit, i.e. some Killing's vectors are degenerated into the lightlike Killing vectors. From the point of view of the second kind phase transitions a cusp point ($p = q = 0$) here is a phase transition point of Schwarzschild's [10] or NUT's gravitational fields from {\it{\bfseries D}} type into {\it{\bfseries N}} type and Kerr's gravitational field from {\it{\bfseries D}} type into {\it{\bfseries III}} or {\it{\bfseries N}} types with the change of the gravitational field symmetry. This is a transition from one "phase" of the gravitational field to another. The parameter $\,p\,$ plays a role of a temperature. The derivative $\,\partial V/\partial p\,$ plays the role of an entropy and $\,\partial^2 V/\partial p^2\,$ corresponds to a thermal capacity.

\section{Summary}

In this paper some results of the lightlike limits of the Schwarzschild, Kerr and NUT solutions on the Weyl's matrixes level and the  investigations connected with the theory of catastrophe are summarized. Under such lightlike limiting procedure the algebraic type of the gravitational field is changed, i.e. the original symmetry of the gravitational field is broken. The lightlike procedure leads to two lightlike particles: the {\it lighton} (the scalar particle) and the {\it helixon} (a spinning particle with the helicity).

On the other hand such change of the algebraic types can be described as the phase transition of second type on the level of Weyl's matrixes, i.e. the phase transition on the level of a curvature of space-time. Hence this phase transition can be named the gravitational phase transition. 

Thus the describing lightlike procedure for Schwarzschild's, NUT's and Kerr's solutions leads to the new metrics of the gravitational wave fields of the lightlike massless particles (the lightons and the helixons) with the total energy and the helicity only. The another physical parameters are lost under such limiting process. It can be said the lightlike sources "have no hairs".

The superposition of such lightlike sources makes possible a construction of the lightlike plane and also of the infinite lightlike pencil (with-out helicity) and of the lightlike ray (with helicity).

\end{document}